\documentclass[%
  journal=jcisd8
]{achemso}

\usepackage[utf8]{inputenc}
\usepackage[T1]{fontenc}
\usepackage{xcolor}
\usepackage[%
  colorlinks=true,
  allcolors=blue
]{hyperref}
\usepackage{xurl}
\usepackage{amsfonts}
\usepackage{amssymb}
\usepackage{amsmath}
\usepackage{mathtools}
\usepackage[ruled,vlined]{algorithm2e}
\usepackage{lmodern}
\usepackage[normalem]{ulem}
\usepackage{sectsty}
\sectionfont{\large}

\DeclarePairedDelimiter\ppar{(}{)}              
\DeclarePairedDelimiter\pnrm{\lVert}{\rVert}    
\DeclarePairedDelimiter\pset{\{}{\}}            

\newcommand{\rfig}[1]{Fig.~\ref{#1}}

\newcommand{\rref}[1]{Ref.~\citenum{#1}}

\newcommand{\bz}{\mathbf{z}}
\newcommand{\bx}{\mathbf{x}}

\title{\large \texttt{NeuralTSNE}: A Python Package for the Dimensionality Reduction of Molecular Dynamics Data Using Neural Networks}

\author{Patryk Tajs}
\author{Mateusz Skarupski}
\author{Jakub Rydzewski}
\email{jr@fizyka.umk.pl}
\affiliation{%
  Institute of Physics,
  Faculty of Physics, Astronomy and Informatics,
  Nicolaus Copernicus University,
  Grudzi\k{a}dzka 5, 87-100 Toru\'n, Poland
}

\begin{document}

\begin{tocentry}
  \begin{center}
    \includegraphics{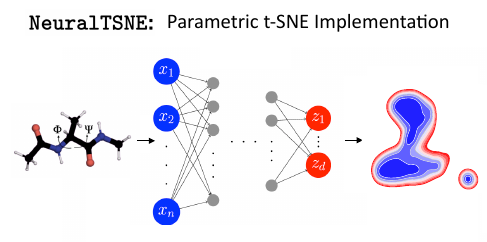}
  \end{center}
\end{tocentry}

\begin{abstract}
Unsupervised machine learning has recently gained much attention in the field of molecular dynamics (MD). Particularly, dimensionality reduction techniques have been regularly employed to analyze large volumes of high-dimensional MD data to gain insight into hidden information encoded in MD trajectories. Among many such techniques, t-distributed stochastic neighbor embedding (t-SNE) is particularly popular. A parametric version of t-SNE that employs neural networks is less commonly known, yet it has demonstrated superior performance in dimensionality reduction compared to the standard implementation. Here, we present a Python package called \texttt{NeuralTSNE} with our implementation of parametric t-SNE. The implementation is done using the PyTorch library and the PyTorch Lightning framework and can be imported as a module or used from the command line. We show that \texttt{NeuralTSNE} offers an easy-to-use tool for the analysis of MD data.
\end{abstract}

\maketitle

\newpage

\section{Introduction}
Understanding molecular dynamics (MD) trajectories depends on our ability to recognize patterns in a high-dimensional representation spanned by features~\cite{ceriotti2019unsupervised,glielmo2021unsupervised,chen2021collective,rydzewski2023manifold,gokdemir2025machine}. Clearly, without prior knowledge about the system at hand, such analysis can be unsystematic and prone to errors. Due to recent implementations of many libraries~\cite{scikit-learn,paszke2019pytorch}, using machine learning (ML) techniques has become relatively straightforward and readily available for applications and further development. Unsupervised ML, particularly dimensionality reduction algorithms, has been especially popular in the field of MD~\cite{molgedey1994separation,wiskott2002slow,coifman2005geometric,singer2009detecting,ceriotti2011simplifying,tiwary2016spectral,wehmeyer2018time,zhang2018unfolding,mardt2018vampnets,chen2019nonlinear,morishita2021time,bonati2021deep,rydzewski2023selecting,rydzewski2023spectral,wang2021state,chen2023discovering}. In brief, these algorithms construct a low-dimensional representation consisting of a few reduced variables that are easier to understand. These variables are referred to as reaction coordinates or collective variables~\cite{rogal2021reaction,noe2017collective}. For MD data, the reduced representation should encode the essential macroscopic characteristics of the process. Much development in recent years has been dedicated to implementing data-driven methods to extract physical information from MD simulations.

One such unsupervised ML method called t-distributed stochastic neighbor embedding~\cite{maaten2008visualizing} (abbreviated as t-SNE) has seen many applications and developments, including to MD data~\cite{rydzewski2016machine,zhou2018t,tribello2019using,zhang2018unfolding,rydzewski2021multiscale,rydzewski2022reweighted,rydzewski2023manifold,appadurai2023clustering}. The t-SNE technique was proposed by van der Maaten and Hinton as an improvement over SNE, which had been developed several years prior by Hinton and Roweis~\cite{hinton2002stochastic}. Following this, inspired by the work on autoencoders~\cite{hinton2006reducing}, van der Maaten published his seminal work on parametric t-SNE~\cite{maaten2009learning} that performed dimensionality reduction using feedforward neural networks (NNs). In the field of MD, the development of parametric t-SNE inspired many techniques, including methods such as stochastic kinetic embedding~\cite{zhang2018unfolding} and multiscale reweighted stochastic neighbor embedding~\cite{rydzewski2021multiscale,rydzewski2022reweighted}. Although the standard version of t-SNE is available in several libraries~\cite{scikit-learn,Policar2024}, there is no general implementation of parametric t-SNE that can be easily used and extended. To this end, we present a Python package called \texttt{NeuralTSNE} with our implementation of the parametric version of t-SNE that employs an NN for dimensionality reduction. Our implementation is done using the PyTorch library~\cite{paszke2019pytorch} and the PyTorch Lightning framework and can be imported as a module or used from the command line. Although here we present several examples of using \texttt{NeuralTSNE} to analyze MD data, the package is general and can be used for any dataset. \texttt{NeuralTSNE} offers a practical and easy-to-use tool for the analysis of molecular processes.

\section{Algorithm}
\begin{figure}
  \includegraphics[width=0.5\textwidth]{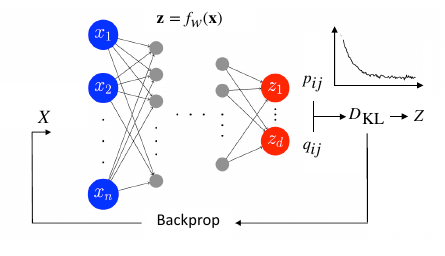}
  \caption{Outline of the learning process implemented in parametric t-SNE. The dataset $X$ is iteratively fed to the NN $f_w$ to map the samples from the feature space to the reduced space. The Kullback-Leibler divergence between the probabilities in feature space $p_{ij}$ and reduced space $q_{ij}$ is minimized to preserve the structure of the data in the reduced space.}
  \label{fig:nn}
\end{figure}

The process of dimensionality reduction in parametric t-SNE, as it is customarily done with training NNs, is iterative and stops when either the maximum number of epochs or a set accuracy is reached. The learning is based on converting pairwise distances between samples in both the feature space and the reduced space into probabilities that measure the similarity of samples. The difference between these probabilities, measured by the Kullback-Leibler divergence ($D_{\mathrm{KL}}$), is minimized during training to preserve the structure of the data represented in the feature space in reduced space~\cite{hinton2002stochastic,maaten2008visualizing} (\rfig{fig:nn}).

In the following, we present the algorithm after van der Maaten~\cite{maaten2009learning}:
\begin{enumerate}
  \item Set perplexity $P$ (the number of effective neighbors) and the number of reduced variables $d$. Define NN $f_w$ with trainable parameters $w$.
  \item Iterate over epochs until convergence is reached:
  \begin{enumerate}
    \item The pairwise distances in the feature space $\pnrm{\bx_i-\bx_j}$ are transformed into conditional probabilities by centering a Gaussian over each sample $\bx_i$ and computing the density of $\bx_j$ under this Gaussian. Thus, the conditional probabilities $p_{j|i}$ and $p_{i|j}$ are given as:
    \begin{equation}
      p_{j|i} = \frac{\exp\ppar*{-\frac{1}{2\sigma_i^2}\pnrm{\bx_i-\bx_j}^2}}{\sum_{i \neq k} \exp\ppar*{-\frac{1}{2\sigma_i^2}\pnrm{\bx_i-\bx_k}^2}},
    \end{equation}
    where $\sigma$ denotes the Gaussian bandwidth. It is set using the bisection method, such that the Shannon entropy of the conditional distribution equals a predefined perplexity, $P_i = 2^{H_i}$, where $H_i = -\sum_j p_{j|i} \log p_{j|i}$ is the Shannon entropy. Then, the conditional probabilities are symmetrized, $p_{ij} = \frac{p_{i|j}+p_{j|i}}{2n}$. (More details on how setting perplexity affects the reduced space can be found in \cite{wattenberg2016use}.)
    \item Calculate the probabilities $q_{ij}$ in the reduced space (dependent on the parameters $w$) based on a t-student kernel:
    \begin{equation}
      q_{ij} = \frac{\ppar*{1+\frac{1}{\alpha}\pnrm{f_w(\bx_k)-f_w(\bx_l))}^2}^{-\frac{\alpha+1}{2}}}{\sum_{i \neq k} \ppar*{1+\frac{1}{\alpha}\pnrm{f_w(\bx_i)-f_w(\bx_k))}^2}^{-\frac{\alpha+1}{2}}},
    \end{equation}
    where $\alpha$ represents the number of degrees of freedom of the distribution (usually set as equal to the dimensionality of the reduced space).
    \item Estimate the loss as the Kullback-Leibler divergence:
    \begin{equation}
      D_{\mathrm{KL}} = \sum_{i \neq j} p_{ij} \log\ppar*{\frac{p_{ij}}{q_{ij}}}
    \end{equation}
    to measure the difference between the probabilities in the feature and reduced spaces.
    \item Update the parameters of NN $w$ by performing backpropagation. The parameters of the NN are adjusted using an optimizer (e.g., Adam~\cite{kingma2014adam}) in such a way that the Kullback-Leibler divergence between the probabilities in the feature and reduced spaces is minimized.
  \end{enumerate}
  \item The trained NN can be used to map the system trajectory from the MD data to the reduced space $Z=\pset{\bz_i}_{i=1}^n$.
\end{enumerate}
The trained NN can also be employed to reduce samples from outside the dataset, in contrast to the standard implementation of t-SNE, which is a considerable advantage for MD datasets. Therefore, the NN trained using parametric t-SNE can be employed in enhanced sampling simulations, where collective variables are needed to bias a simulation~\cite{valsson2016enhancing,bussi2020using,henin2022enhanced}.

Note that the algorithm presented by van der Maaten~\cite{maaten2009learning} also incorporated a weight initialization procedure of the NN using Boltzmann machines~\cite{ackley1985learning}. Our implementation, however, does not include this procedure.

\section{Implementation}
The \texttt{NeuralTSNE} package is implemented in Python 3.11+ and available for all major operating systems via the Python Package Index (PyPI). It uses the PyTorch library~\cite{paszke2019pytorch} for the training of NNs. It can be run on a GPU for faster computations. \texttt{NeuralTSNE} also employs the PyTorch Lightning framework (\url{https://github.com/Lightning-AI/lightning}), which serves as a high-level wrapper for PyTorch to simplify the training process for the users. The PyTorch Lightning framework enables us to incorporate various callbacks into the training process. Moreover, methods for model fitting, or prediction, are named in a manner consistent with the nomenclature used in the scikit-learn library, so they can be accessed using the functions named fit and predict.

The package can be imported as a module or used as a command-line tool \texttt{neural-tsne}. The code is hosted on GitHub (\url{https://github.com/NeuralTSNE/NeuralTSNE}) and licensed under the MIT License. The repository is coupled to CI/CD via GitHub Actions, performing automated testing upon changes or proposed changes to the main branch. The package also uses the pytest and unittest testing frameworks. The documentation can be found at \url{https://NeuralTSNE.github.io/NeuralTSNE}. We provide several examples prepared as Jupyter notebooks containing the calculations for the model system presented in this work, the MNIST dataset, and an example of using the package as a command-line tool.

\section{Examples}
Examples are provided in the form of Jupyter notebooks to provide easy-to-follow tutorials for the users. The MD datasets are generated from 100-ns parallel tempering simulations~\cite{parallel_tempering} of alanine dipeptide and alanine tetrapeptide in vacuum. The simulations were conducted using the Gromacs code~\cite{gromacs}. The alanine dipeptide dataset includes 45 features, which are pairwise distances between heavy atoms. To represent the alanine tetrapeptide, we compute the sines and cosines of the $\Phi$ and $\Psi$ dihedral angles, resulting in a total of 12 features. The features of these systems are calculated from replicas at a temperature of 300 K. Further details about these MD datasets are described in \rref{rydzewski2021multiscale}. They are available for download at Zenodo (DOI: \href{https://doi.org/10.5281/zenodo.4756093}{10.5281/zenodo.4756093}) and from the PLUMED NEST under plumID:21.023 at \url{https://www.plumed-nest.org/eggs/21/023/}.

\begin{figure*}
  \includegraphics{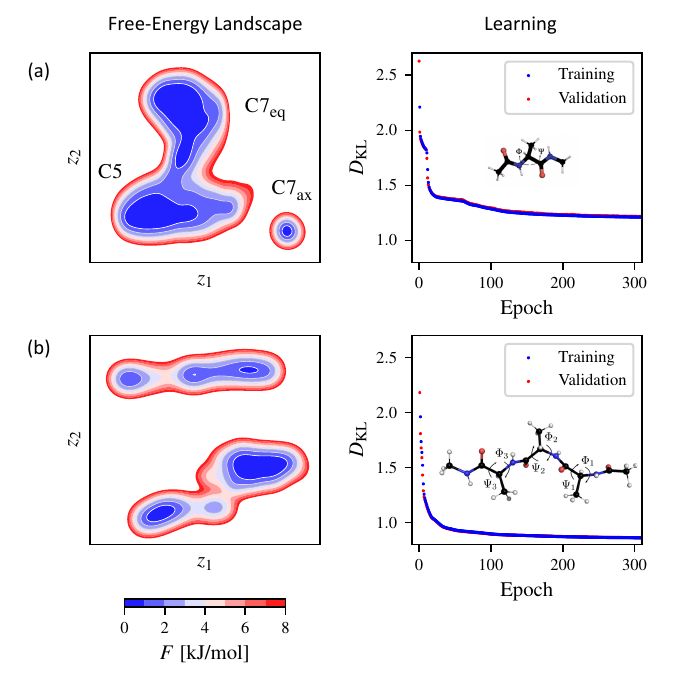}
  \caption{Results of applying parametric t-SNE to the example systems: (a) alanine dipeptide and (b) alanine tetrapeptide. The first column depicts free-energy landscapes calculated from the reduced representation, while the second column shows the training and validation loss obtained during dimensionality reduction.}
  \label{fig:example}
\end{figure*}

In both examples, we employ NNs with two hidden layers and ReLU activation functions. Each layer contains a number of nodes equal to 75\% of the input size. The output of the NNs comprises two variables. We employ batches of size 1000 and set the perplexity to 30. The datasets are divided into training and validation sets, with the validation set representing 20\% of the full dataset. We conduct the learning over 500 epochs, but we implement early stopping to terminate the process when the validation loss reaches a precision of $10^{-6}$. As an optimizer, we use Adam with a learning rate of $10^{-3}$ and default parameters~\cite{kingma2014adam}. The detailed process of using parametric t-SNE for these MD datasets is shown in the examples directory of the package.

Our results are depicted in \rfig{fig:example}. In the first column, we present free-energy landscapes calculated in a reduced space learned through parametric t-SNE. The second column shows the minimization of the Kullback-Leibler divergence. We observe that the loss reaches a plateau after 100 epochs in both examples, with the validation loss closely matching the training loss. In both examples, training concludes before 500 epochs, achieving the specified validation loss precision.

The reduced space of alanine dipeptide consists of three metastable states (\rfig{fig:example}a). The states C7$_\mathrm{eq}$ and C5 are separated by a free-energy barrier of around 2 kJ/mol, indicating that the transitions between these states are relatively fast. The transition to the third state, C7$_\mathrm{ax}$, is much slower, as indicated by a barrier of approximately 10 kJ/mol. The free-energy landscape of alanine dipeptide constructed using parametric t-SNE closely matches that calculated for the $\Phi$ and $\Psi$ dihedral angles, which are known to map the most interesting characteristics of the system. The free-energy landscape of alanine tetrapeptide is more complex than that of alanine dipeptide. We can see in \rfig{fig:example}b that parametric t-SNE identifies three major metastable states, some having minor substates. These results closely resemble a reduced space obtained using MRSE in our previous work~\cite{rydzewski2021multiscale}, showing that parametric t-SNE is able to identify the most important metastable states of alanine tetrapeptide. The remaining states are located high in the free-energy landscape and can only be efficiently sampled using enhanced sampling~\cite{rydzewski2021multiscale}.

\section{Conclusions}
In this application note, we present a Python package called \texttt{NeuralTSNE} with our implementation of parametric t-SNE that employs an NN for dimensionality reduction. Our implementation is done using the PyTorch library and the PyTorch Lightning framework and can be imported as a module or used from the command line. The package is designed in a modular way, allowing for integration with many useful modules already available in PyTorch, such as defining neural networks. Although here we present several examples of using \texttt{NeuralTSNE} to analyze MD data, the package is general and can be used for any dataset. \texttt{NeuralTSNE} offers a practical and easy-to-use tool for the analysis of molecular processes. The package can be easily extended to include more spatial unsupervised learning techniques, such as multiscale reweighted stochastic embedding~\cite{rydzewski2021multiscale,rydzewski2022reweighted} or spectral map~\cite{rydzewski2023spectral,rydzewski2024learning,rydzewski2024tse}. For future development, we plan to extend the implementation and integrate NNs trained using our package so that it can be used for biasing in enhanced sampling simulations in PLUMED~\cite{plumed,plumed-nest,plumed-tutorials} with the PyTorch module~\cite{bonati2023unified}. We hope that \texttt{NeuralTSNE} can be useful for the community in the analysis of MD data.

\section{Data Availability}
All the data and PLUMED input files required to reproduce the datasets analyzed in this paper are available on Zenodo (DOI: \url{10.5281/zenodo.4756093}) and PLUMED-NEST, the public repository of the PLUMED consortium~\cite{plumed-nest}, as plumID:21.023. The implementation of the \texttt{NeuralTSNE} package and the examples are available in a git repository (\url{https://github.com/NeuralTSNE/NeuralTSNE}). The documentation can be accessed at \url{https://NeuralTSNE.github.io/NeuralTSNE/}.

\section{Supporting Information}
Supporting Information is available free of charge at \url{https://pubs.acs.org/}.
\begin{itemize}
  \item \texttt{NeuralTSNE} (\href{https://github.com/NeuralTSNE/NeuralTSNE/archive/refs/heads/main.zip}{zip})
\end{itemize}

\subsubsection*{Author Contributions}
Conceptualization: P.~T. and J.~R. Software: P.~T. Data curation: all authors. Formal analysis: M.~S. and J.~R. Methodology: P.~T. and J.~R. Writing-original draft: J.~R. Writing-review and editing: all authors.

\subsubsection*{Funding}
We acknowledge support from the Ministry of Science and Higher Education in Poland and the National Science Center in Poland (Sonata 2021/43/D/ST4/00920, ``Statistical Learning of Slow Collective Variables from Atomistic Simulations'').

\subsubsection*{Notes}
The authors declare no competing financial interest.

\section{Acknowledgements}
We thank Omar Valsson for his feedback on the manuscript.

\bibliography{bib/main.bib}

\end{document}